# Band engineering of Dirac surface states in topological insulators-based van der Waals heterostructures

Cui-Zu Chang, [1,2] Peizhe Tang, [1] Xiao Feng, [1,2] Kang Li, [2] Xu-Cun Ma, [1] Wenhui Duan, [1,3] Ke He, [1] and Qi-Kun Xue [1,2]

[1] State Key Laboratory of Low-Dimensional Quantum Physics, Department of Physics, Tsinghua University, Beijing 100084, China

[2] Beijing National Laboratory for Condensed Matter Physics, Institute of Physics, Chinese Academy of Sciences, Beijing 100190, China

[3] Institute of Advanced Study, Tsinghua University, Beijing 100084, China

Corresponding authors: czchang@mit.edu(C. Z. C), dwh@phys.tsinghua.edu.cn (W. H. D) and kehe@mail.tsinghua.edu.cn(K. H.)

The existence of gapless Dirac surface band of a three dimensional (3D) topological insulator (TI) is guaranteed by the non-trivial topological character of the bulk band, yet the surface band dispersion is mainly determined by the environment near the surface. In this *Letter*, through *in-situ* angle-resolved photoemission spectroscopy (ARPES) and the first-principles calculation on 3D TI-based van der Waals heterostructures, we demonstrate that one can engineer the surface band structures of 3D TIs by surface modifications without destroying their topological non-trivial property. The result provides an accessible method to independently control the surface and bulk electronic structures of 3D TIs, and sheds lights in designing artificial topological materials for electronic and spintronic purposes.

A topological insulator (TI) has conducting surface states (SSs) protected by the non-trivial topological property of its insulating bulk bands [1, 2]. The topological SSs of TIs have Dirac-cone-shaped band structure and are spin-momentum locked [3-6]. Exotic physical phenomena such as Majorana Fermions[2] and quantum anomalous Hall effect [7-9] have been proposed based on TIs, which brings the materials tremendous application potential in future spintronics and fault-tolerant quantum computation. $Bi_2Se_3$, $Bi_2Te_3$ and $Sb_2Te_3$, are most-studied 3D TIs by virtue of their simple composition and single Dirac cone band structure [3-6]. To realize various quantum phenomena and applications of TIs, one needs to be able to engineer the band structures of Dirac SSs (DSSs) while keeping bulk insulating. However, for a homogeneous TI material its bulk and surface properties are interdependent; the efforts of optimizing one will usually deteriorate the other. For example, to tune the Fermi level of $Sb_2Te_3$ into bulk gap, one can dope Bi in it. This would inevitably push the Dirac point (DP) below the valence band maximum, burying its exotic electronic properties [10, 11]. Thus flexible and stable tuning of the properties of SSs and bulk bands of 3D TIs is a prerequisite for obtaining appropriate TIs for different purposes.

In fact, the unique character of TI provides a solution to the problem [1, 2]. Although the existence of the DSSs of a TI is guaranteed by the topological character of the bulk band, the specific band structure of the DSSs is mainly determined by potential landscape near the surface. Thus just like conventional SSs, the band structure of topological DSS can be modulated by surface modifications but will still be protected as long as time reversal symmetry is kept [12].

Inspired by this idea, we use molecular beam epitaxy (MBE) to deposit different van de Waals cover layers (*e.g.*, one quintuple layer (QL) TI film (hereafter referred as $TI_c$) or one bilayer (BL) Bi (111)) on the another kind 3D TI films (hereafter referred as $TI_b$), as shown in **Fig.1**. Combined with angle-resolved photoemission spectroscopy (ARPES) measurements and the density functional theory (DFT) calculations, we find that the SS bands can be engineered by heteroepitaxy of the other material. For the hetero-structures deposited by one QL $TI_c$, the energy dispersion of the SS and the surface carrier type are determined by the cover layer. While, for $Bi_2Se_3$ thin films covered by one BL Bi (111), the band crossing of the spin-polarized states originated from the large Rashba splitting of Bi BL coexists with the DSS of $Bi_2Se_3$; and owing to the strong surface hybridization, the energy dispersion and Fermi surface (FS) of DSS have been strongly modified. This work elucidates that depositing heterogeneous cover layers on TI thin films is a feasible approach to engineer the topological DSS, which allows us to achieve ideal TIs.

Single crystal $TI_b$ films were grown on graphene terminated 6H-silicon carbide (0001) substrate [13]. When a thick $TI_b$ films is grown, 1QL $TI_c$ was deposited onto it. During the growth of 1QL $TI_c$, the substrate temperature is slightly lower (~10℃) than the growth temperature of thick $TI_b$ films to minimize the possible intermixing. For 1BL Bi (111) on 20QL $Bi_2Se_3$ heterostructure, 1BL Bi (111) film is deposited on the thick $TI_b$ films at room temperature and annealed at 200℃ for one hour [14].

**1QL $TI_c$ film on $TI_b$**: **Figure 2 (a)** show the band dispersions and corresponding MDC along the $\bar{\Gamma}$-$\bar{K}$ direction for 10QL $Bi_2Se_3$ films measured at $T$~150K, respectively. Topological DSS and quantum well states of the conduction band are clearly observed and the DP is located at ~395meV below the Fermi level ($E_F$), much lower than that of the thick $Bi_2Se_3$ films (**Fig.4 (a)**). As previously reported for 1 QL $Bi_2Te_3$ film grown on Si (111) substrate, the SSs open a gap at the DP position [15]. However, when 1QL $Bi_2Te_3$ is on top of 10QL $Bi_2Se_3$ films, a single gapless Dirac cone is observed in the bulk band gap. The surface band dispersion near DP is almost identical as the DSS of slightly hole doped 3D TI $Bi_2Te_3$ films [16], but the DP is faintly lifted out of the bulk valence bands, as shown in **Figs. 2(b).** The significant change of DSS band dispersion from that of $Bi_2Se_3$ to that of $Bi_2Te_3$, instead of coexistence, indicates that 1QL $Bi_2Te_3$ basically covers the whole surface of 10QL $Bi_2Se_3$. This ultrathin 1QL $Bi_2Te_3$ changes the DSS of 10QL $Bi_2Se_3$, and makes the 1QL $Bi_2Te_3$/10QL $Bi_2Se_3$ heterostructure show DSS of $Bi_2Te_3$ characteristic. This observation can be understood by TI band theory[1, 2], where 3D $Bi_2Te_3$ shares the same bulk non-trivial $Z_2$ number with 3D $Bi_2Se_3$, while 1 QL $Bi_2Te_3$ cover layer does not change the bulk non-trivial $Z_2$ topology of 10QL $Bi_2Se_3$, resulting in a DSS which is mainly determined by the surface Hamiltonian. **Figs.2(c)** are the band dispersions and the corresponding MDCs along the $\bar{\Gamma}$-$\bar{K}$ direction for 10QL $Sb_2Te_3$, respectively. DSS is clearly observed and DP is located at +45meV above the $E_F$, indicating the *p*-type carriers. When 1QL $Bi_2Te_3$ is covered on 10QL $Sb_2Te_3$, a single gapless Dirac cone is observed in the bulk band gap below the $E_F$, while the surface band dispersion near DP is similar with DSS of hole doped 3D TI $Bi_2Te_3$ films [15], as shown in **Figs. 2(d)**. This prominent DSS induced by 1QL $Bi_2Te_3$ from $Sb_2Te_3$ to $Bi_2Te_3$ is also reasonable due to their same bulk $Z_2$ topology.

To further confirm the ARPES results, we carried out the *ab initio* DFT calculations to study the electronic structures of 1QL $TI_c$/6QL $TI_b$ heterostructure, such as 1QL $Sb_2Te_3$ on 6QL $Bi_2Te_3$ thin films and vice versa. The calculated band structures are shown in **Fig.3**, where the red shows the contributions from the top cover layer. Compared with the pristine 6QL $Bi_2Te_3$ thin film (**Fig. 3(a)**), the DSS of the heterostructure still exists and mainly contributed by covered 1QL $Sb_2Te_3$ thin film, showing the characteristic of $Sb_2Te_3$ SS (**Fig. 3(b)**), meanwhile the quantum well states contributed by $Bi_2Te_3$ also retain in this heterostructure. Similar phenomena are also observed in 1QL $Bi_2Te_3$ on 6QL $Sb_2Te_3$ thin films, whose DSS has been engineered by the top cover layer only 1QL $Bi_2Te_3$ (**Figs. 3(c) and (d)**), consistent with the experimental observation (**Figs. 2(c) and (d)**). These results suggest that the heterogeneous growth of other hetero-type $TI_c$ that also belongs to $Bi_2Se_3$ family cannot change the bulk band non-trivial $Z_2$ topology, while can efficiently modulate the DSS. This artificial band engineering approach guides us to design an ideal TI heterostructure with maintenance of high surface mobility and simultaneously tunable carrier type but insulating bulk states as an excellent platform for the further study.

**1BL Bi (111) on $Bi_2Se_3$**: Besides the heterostructure between 1QL $TI_c$ and 3D $TI_b$, 1BL Bi (111) film have also been tried to deposit on 3D TI films to form heterostructure. The electronic band dispersions of pristine and 1BL Bi (111) covered 20QL $Bi_2Se_3$ thin films measured by *in situ* ARPES along the $\bar{\Gamma}$-$\bar{K}$ direction are shown in **Figs. 4(a)** and **(b)**, respectively. Two bands with nearly linear dispersion cross each other at $\bar{\Gamma}$ point, forming a Dirac cone in pristine 20QL $Bi_2Se_3$ thin films. The bulk conduction band can be observed and the DP is embedded in -190meV below the $E_F$. Intriguingly, this situation will be changed when 1BL Bi (111) is deposited on surface. Besides the topological DSSs of 20QL $Bi_2Se_3$, an additional pair of large Rashba-type splitting bands dispersing downwards from $\pm 0.08 Å^{-1}$ at FS is raised. The brighter two bands crossing state indicate the DP of the bottom 20QL $Bi_2Se_3$, which is located at -300meV below the $E_F$, it indicates that 1 BL Bi (111) introduces some electrons into 20 QL $Bi_2Se_3$. Additional bright state feature crossed by the Rashba-type sub-bands emerge in the vicinity of top cone of DSS [16-18]. Due to BL Bi (111), there is a distinct modification for the band dispersions of DSS, which is extensively analyzed in detail in the corresponding MDC. For the constant energy contours of bands in 1BL Bi (111)/ 20 QL $Bi_2Se_3$, shown in **Fig.4 (c)**, the strong hexagonal warping effect can be observed in the FS of the outermost band, which shows snowflake shape and nest together with the FS of TI's surface state, as well as a slight warping effect. On the contrary, the inner FS of the Rashba-split sub-bands shows nearly circular shape, which has a Rashba-like non-degenerate spin texture [19].

1BL Bi (111) layer changes the band dispersion of DSS contributed by the substrate of $Bi_2Se_3$, the most prominent feature is a tunability of DSS slope near DP, in other words, the Fermi group velocity of Dirac Fermions near DP is changed. Now the binding energy referred to the position of DP in each DSS of 20QL $Bi_2Se_3$ at different certain momentum vectors are plotted and fitted using a proportional function in **Fig. 4(d)**. The slope of the fitting line indicates the Fermi group velocity of the Dirac fermion of each DSS. The fitted Fermi group velocity of DSS in 1BL Bi (111)/ 20 QL $Bi_2Se_3$ is $v_F \sim 3.24 \times 10^5$ m/s, but that in 20QL $Bi_2Se_3$ is $v_F \sim 5.09 \times 10^5$ m/s. 1BL Bi(111) on surface provides additional degree of freedom to engineer DSS in TIs.

In order to further understand the physical origin of the unusual band structures, we performed the scanning tunneling microscopy (STM) measurements and DFT calculations to identify the configuration of $Bi_2Se_3$ thin film covered by 1BL Bi (111). By delicate control of the Bi (111) growth condition, we were able to achieve BL-by-BL growth [14]. **Figs. 5(a)** and **(b)** show the STM images of 1BL Bi (111) and 0.5BL Bi (111) on 20QL $Bi_2Se_3$, respectively. Atomically flat morphology and sharp 1×1 reflection high-energy electron diffraction (RHEED) patterns [19] demonstrate BL-by-BL growth mode and the high crystal quality of BL Bi (111), and Moiré patterns with the period ~4.57nm can be seen in high resolution STM images (**Fig. 5 (b)**), which originates from the in-plane lattice constant mismatch between Bi (111) (~4.54Å) and $Bi_2Se_3$ (~4.13Å). The simulated geometric structure of this heterostructure is shown in **Fig. 5(c)**, where Moiré patterns can also be observed. We choose three typical configurations (hereafter simplified as $AB^1$, $AB^2$ and AA stacking respectively) as prototypes to study their corresponding electronic structures via DFT method, whose calculated electronic structures are shown in **Figs. 5(d)-(f)**. By compared with one another, we can find some common features in all three configurations: (1) a large Rashba splitting can be observed for the bands which are mainly contributed by the inserted 1BL Bi(111) layer and hybridize with bulk states of substrate along $\bar{\Gamma}$-$\bar{M}$ and $\bar{\Gamma}$-$\bar{K}$ directions; (2) due to the inversion symmetry breaking, a Dirac cone ($D_r$ shown in **Figs. 5(d)-(f)**) from Rashba splitting of one BL Bi(111) coexists with the DSS of $Bi_2Se_3$ ($D_s$ shown in **Figs. 5(d)-(f)**), these results are consistent with the experiments and the previous report [16-18]; (3) from the Bader analysis [27], it is found that some electrons transfer from the covered BL Bi(111) layer to $Bi_2Se_3$ thin films, and just like the observation *in situ*

ARPES the sample is electron doped; (4) in contrast to the un-hybridized DSS located on the bottom surface, the DSS on top surface is modified with reduced Fermi group velocity, which originates from the strong hybridization between the Bi(111) BL and top DSS of $Bi_2Se_3$ thin film.

Another interesting discovery on 1BL Bi (111)/20QL $Bi_2Se_3$ heterostructure is the FSs via ARPES. To provide the insight, we analyze the contribution of the calculated states around the DSS on the top surface (dashed line in **Fig 5.(f)**), and plot their calculated FSs and related spin-textures on experimental FSs in **Fig. 4(c)**. As shown in **Fig. 5(g)**, three calculated Bloch states in FSs are consistent with the experimental observation. The inner red one originates from the large Rashba splitting of Bi (111) BL, because the plotted FS is under $D_r$, the helicity of the spin-texture is right-handed [18]. The two cyan Bloch states are mainly contributed by the top DSS of $Bi_2Se_3$ film, the inner circle is the electron-like state with left-handed helicity spin-texture; and the outer hexagonal one is the hole-like state with right-handed helicity spin-texture but its energy dispersion has been strongly modified by the surface hybridization which also enhances its hexagonal warping effect (**Fig. 5(g)**). The experimental FS of Rashba-like sub-bands is above $D_r$ (see in **Fig. 4(c)**), and the helicity of the spin-texture should be left-handed, and therefore, an alternating right-left-left helicity for the spin of consecutive FS of Rashba-like sub-bands and DSS is introduced, as shown in top right corner of **Fig. 5(g)**. Owing to the similar spin-momentum locking behavior between the red circular and the hexagonal Bloch states, we expect that the spin-dependent charge density wave could be observed in this system.

In summary, combined with *in situ* ARPES experiments and DFT calculations, we demonstrate that the DSS of 3D TI could be artificially engineered via van der Waals heterostructure growth of the cover layers. For the heterostructures deposited by one QL $TI_c$, the thin films is still topologically nontrivial, but the properties of DSS, such as band dispersion and surface carrier type, can be engineered through the top cover layers. Moreover, because of the change of surface carrier type from the top to the bottom, a topological *p-n* junction naturally formed along the horizontal direction of the heterostructure, such pair of *p-* and *n*-type on opposite surfaces is likely to give rise to a topological exciton condensate with novel properties [28]. In addition, for 1 BL Bi (111) on $Bi_2Se_3$ thin films, the strong surface hybridization will not influence the coexistence of two DPs, one is from the strong Rashba splitting of topmost Bi BL and the other is DSS of $Bi_2Se_3$ top surface, but can re-construct their spin-polarized FS. This work provides a method for artificially and independently engineering the DSS, and paves the way to design new TI materials for future spintronics and quantum computations.

We thank J. S. Moodera and M. D. Li for helpful discussions, and S. L. He and X. J. Zhou for the help in the generation of FS. This work was supported by the National Natural Science Foundation of China, the ministry of Science and Technology of China, and the Chinese Academy of Sciences.

# Figures and Figures captions

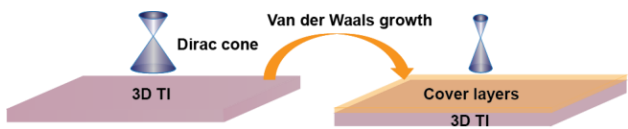

**Fig.1** (color online). Schematics of the van der Waals growth of cover layers on 3D topological insulator thin films.

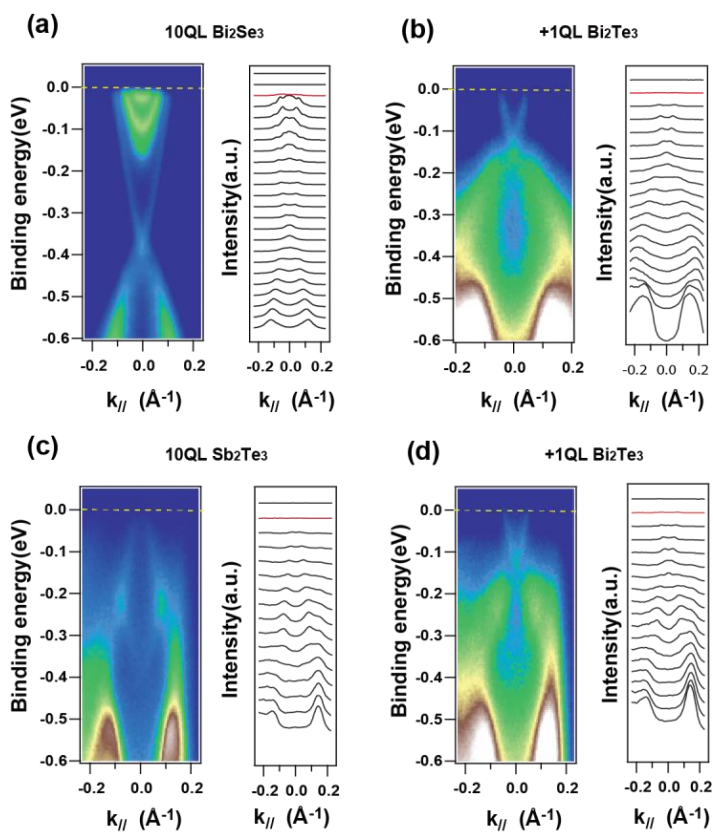

**Fig.2:** (color online). Dirac surface states engineering in 3D TI films when another 1QL TI film is deposited on the surface. ARPES band structures and corresponding momentum distribution

curves (MDCs) of (a) 10QL $Bi_2Se_3$, (b) 1QL $Bi_2Te_3$/10QL $Bi_2Se_3$, (c) 10QL $Sb_2Te_3$, and (d) 1QL $Bi_2Te_3$/10QL $Sb_2Te_3$. All the spectra were taken along $\bar{\Gamma}$-$\bar{K}$ direction at $T\sim150K$.

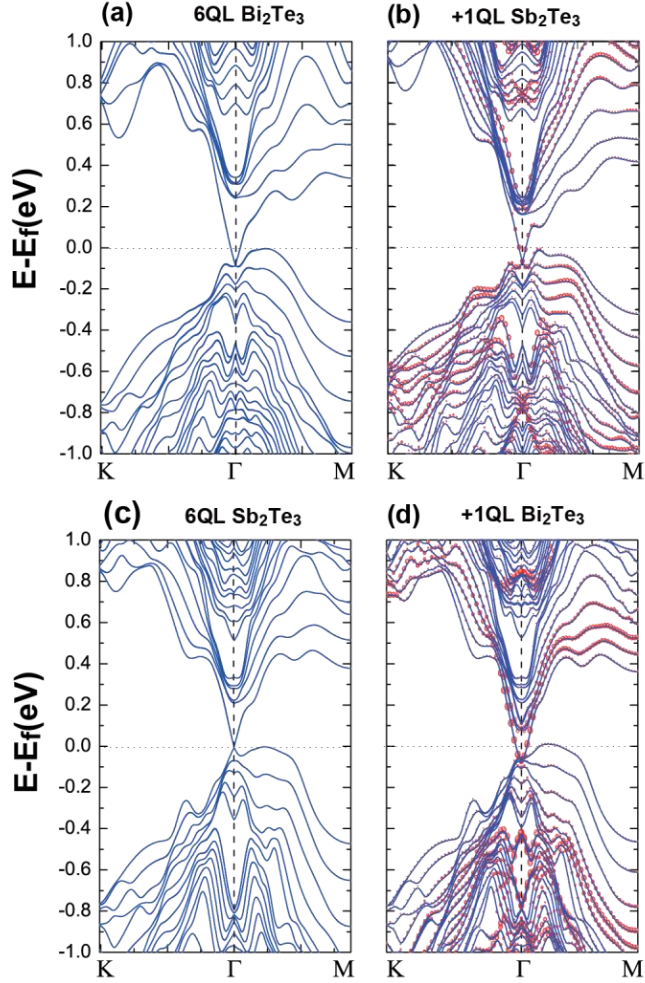

**Fig.3** (color online). First principle calculations for the artificially engineered Dirac surface states. The DFT band structures of (a) 6QL $Bi_2Te_3$, (b) 1QL $Sb_2Te_3$/6QL $Bi_2Te_3$, (c) 6QL $Sb_2Te_3$ and (d) 1QL $Bi_2Te_3$/6QL $Sb_2Te_3$. The red solid and hollow dots in (b) and (d) show the bands induced by the topmost layer 1QL $Sb_2Te_3$ or $Bi_2Te_3$.

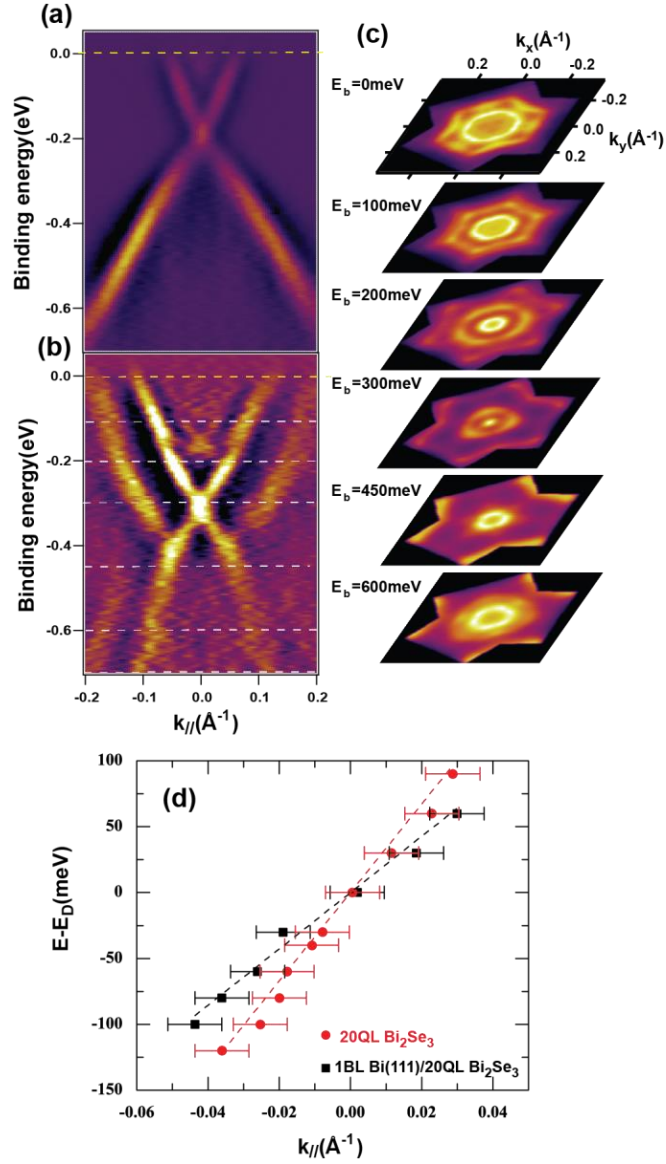

**Fig.4** (color online). (a) ARPES band map of 20QL $Bi_2Se_3$. (b) ARPES band map and (c) Constant energy contours of the Dirac surface states of 1BL Bi (111)/20QL $Bi_2Se_3$ hetero-structure. All the spectra were taken along $\bar{\Gamma}$-$\bar{K}$ direction at room temperature. (d)

Comparison of Dirac surface states momentum vector of 20QL $Bi_2Se_3$ and 1BL Bi (111)/ 20QL $Bi_2Se_3$. The dashed line is the fitting curve, whose slope indicates Fermi group velocity of Dirac surface states in 20QL $Bi_2Se_3$ and 1BL Bi (111)/20QL $Bi_2Se_3$.

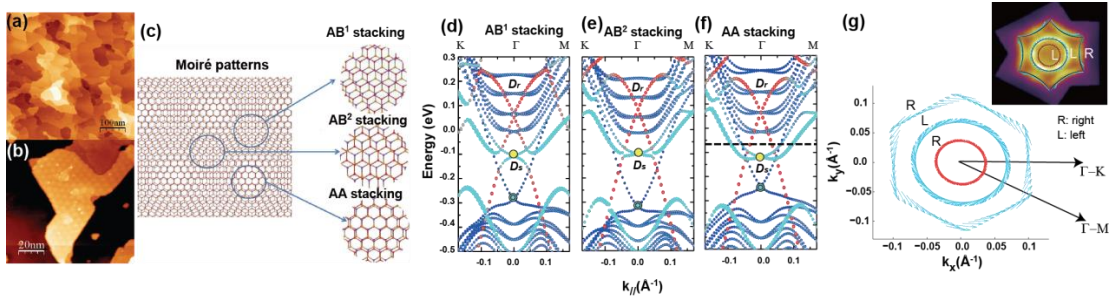

**Fig.5** (color online). (a) STM images of 500nm×500nm taken from 1BL Bi (111)/ 20QL $Bi_2Se_3$ film. (b) High resolution STM images of 100nm×100nm taken from 0.5BL Bi (111)/ 20QL $Bi_2Se_3$ film, Moiré patterns can be observed. (c)Simulated Moiré patterns for 1BL Bi (111) on top of $Bi_2Se_3$ films, and the schematic diagrams of stacking pattern are shown in the left, including $AB^1$ stacking, $AB^2$ stacking and AA stacking. (d-f) The corresponding calculated band structures. Fermi levels are set to zero. The top and bottom DSS of 6QL $Bi_2Se_3$ are marked by the yellow and blue dots, respectively. The red and cyan solid dots in the band structures stand for the contribution of the top 1BL Bi (111) and topmost one QL of $Bi_2Se_3$, respectively. (g) In-plane spin-momentum relationship of the constant energy contour whose energy is marked by the dashed line in (d) and spin texture of the experimental Fermi surface in 1BL Bi (111)/20QL $Bi_2Se_3$. The arrows indicate the spin direction, R means right-handed helicity, L means left-handed helicity.

# Supplementary Material

# Band engineering of Dirac surface states in topological insulators-based van der Waals heterostructures

Cui-Zu Chang,[1,2] Pei-Zhe Tang,[2] Xiao Feng,[1,2] Kang Li,[1] Xucun Ma,[1,2] Wen-Hui Duan,[2,3] Ke He,[1,2] and Qi-Kun Xue[2,1]

[1] Beijing National Laboratory for Condensed Matter Physics, Institute of Physics, Chinese Academy of Sciences, Beijing 100190, China

[2] State Key Laboratory of Low-Dimensional Quantum Physics, Department of Physics, Tsinghua University, Beijing 100084, China

[3] Institute of Advanced Study, Tsinghua University, Beijing 100084, China

Corresponding authors: czchang@mit.edu(C. Z. C), dwh@phys.tsinghua.edu.cn (W. H. D) and kehe@iphy.ac.cn(K. H.)

Ⅰ. **The information of ARPES measurements and the method of DFT calculations**

*ARPES measurements*

The ARPES measurements were carried out *in situ* in an ultra-high vacuum (UHV) chamber (base pressure $<2\times10^{-10}$ mbar) equipped with a Scienta SES2002 analyzer with a resolution of 15 meV and a Gammadata He discharging lamp as photon source (He-Iα, 21.21eV).

*DFT calculations*

For DFT calculations, we use the Vienna *ab initio* simulation package [20] to simulate the electronic properties with the projector augmented wave potential [21] and the Perdew-Burke-Ernzerhof generalized gradient approximation (GGA-PBE) functional [22,23]. A slab model with the in-plane lattice constants from experimental values [24] is initially constructed, which contains the 6QL original TI$_b$, the cover layer and a vacuum layer of 15 Å along the *z* direction. And the cover layer is strained to match the substrate lattice parameter. During the structural relaxation, the topmost four atomic layers of TI$_b$ and the cover layer are allowed to fully relax until the residual forces are less than 0.01eV/ Å, while the dipole correction is introduced to avoid spurious interaction between periodic images of the slabs [25]. The cutoff energy of the plane wave basis set is 340 eV and $13\times13\times1$ Monkhorst-Pack *k* points are used to ensure convergence. Spin-orbit coupling is included in calculating the electronic structures [26].

## Ⅱ. ARPES band maps of 10QL Bi$_2$Se$_3$, Bi$_2$Te$_3$ and Sb$_2$Te$_3$ films

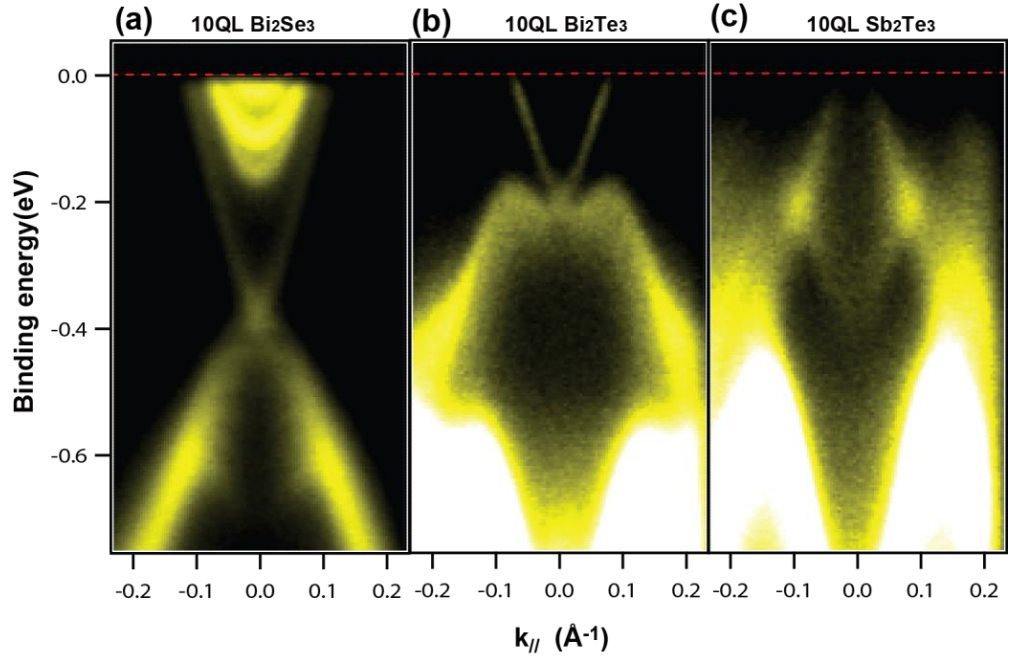

**Fig. S1.** ARPES band maps of 10QL Bi$_2$Se$_3$ (a), 10QL Bi$_2$Te$_3$ (b), 10QL Sb$_2$Te$_3$ (c) grown on graphene terminated 6H-SiC (0001). All the spectra were taken along $\bar{\Gamma}$-$\bar{K}$ direction at $T$~150K.

## Ⅲ. RHEED patterns of the films

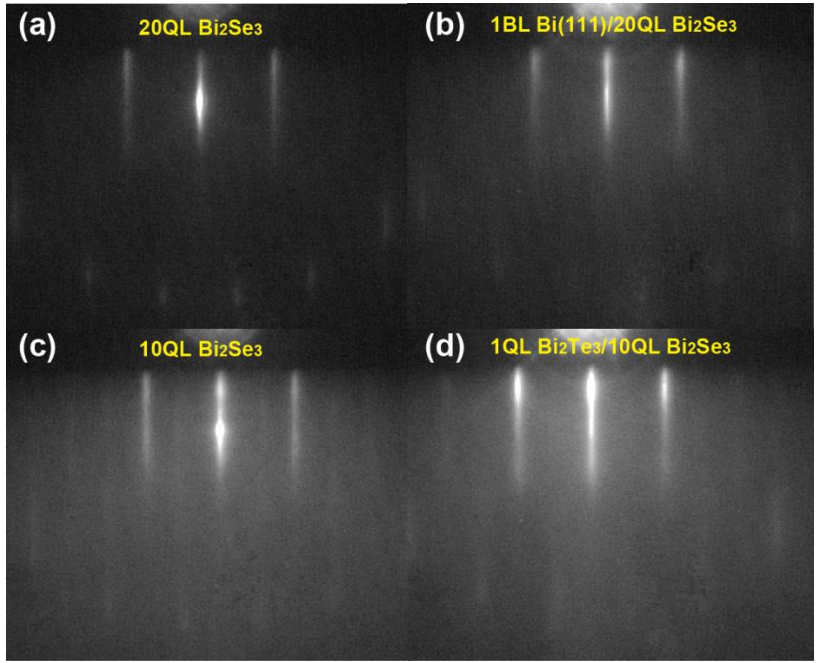

**Fig. S2.** RHEED patterns of 20QL $Bi_2Se_3$ (a), 1BL Bi (111)/$Bi_2Se_3$ hetero-structure (b), 10QL $Bi_2Se_3$ (c) and 1QL $Bi_2Te_3$/10QL $Bi_2Se_3$ hetero-structure grown on graphene terminated 6H-SiC (0001). Sharp 1×1 diffraction streaks demonstrate the high crystalline quality of these films.

# Ⅳ. Dirac surface states engineering in 3D TI films when another 1QL Bi$_2$Se$_3$ film is deposited on the surface

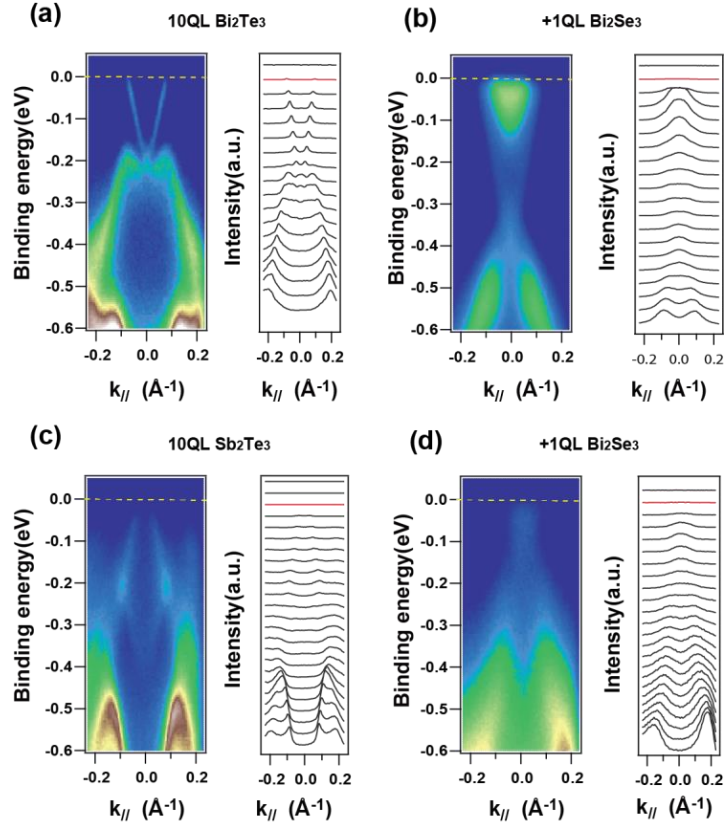

**Fig. S3** ARPES band structures and corresponding momentum distribution curves (MDCs) of (a) 10QL Bi$_2$Te$_3$, (b) 1QL Bi$_2$Se$_3$/10QL Bi$_2$Te$_3$, (c) 10QL Sb$_2$Te$_3$, and (d) 1QL Bi$_2$Se$_3$/10QL Sb$_2$Te$_3$. All the spectra were taken along $\bar{\Gamma}$-$\bar{K}$ direction at $T$~150K.

# Ⅴ. The calculations for in-plane spin component of the outer two bands in one constant energy contour of 1BL Bi (111)/6QL Bi$_2$Se$_3$

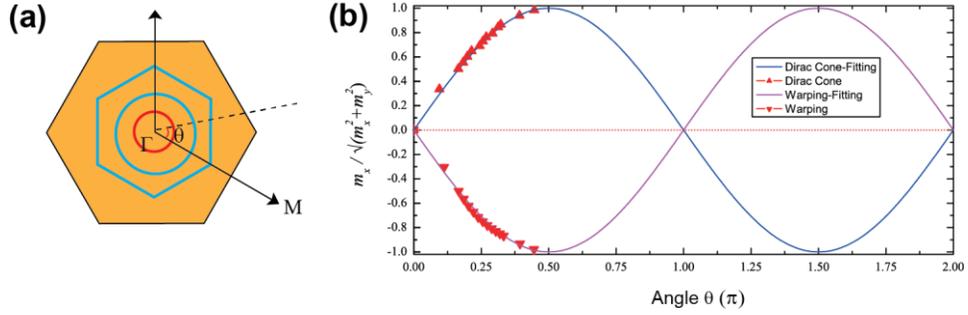

**Fig. S4** (a) The schematic diagram for the first 2D Brillouin zone of the projected (111) surface in 1BL Bi (111)/6QL Bi$_2$Se$_3$ hetero-structure. (b) The calculated in plane spin component and the fitting curves of the outer two bands in one constant energy contour of 1BL Bi (111)/ 6QL Bi$_2$Se$_3$ hetero-structure. The energy of the constant energy contour is marked by the dashed line in Fig.5 (f).